\documentclass[10pt,twocolumn,letterpaper]{article}

 \usepackage{cvpr}              

\usepackage[dvipsnames]{xcolor}
\usepackage{graphicx}
\usepackage{afterpage} 
\usepackage{amsmath}
\usepackage{amssymb}
\usepackage{booktabs}
\usepackage{lineno}
\usepackage{enumitem}
\usepackage{amsmath}
\usepackage{amssymb}
\usepackage{natbib}
\usepackage{commath}
\usepackage{adjustbox}
\usepackage{geometry} 
\usepackage{tabularx}
\usepackage{caption}
\usepackage{float}
\usepackage{verbatim}
\usepackage{multicol}
\usepackage{booktabs}
\usepackage{multirow}

\definecolor{cvprblue}{rgb}{0.21,0.49,0.74}
\usepackage[pagebackref,breaklinks,colorlinks,citecolor=cvprblue]{hyperref}

\def\paperID{40} 
\def\confName{CVPR}
\def\confYear{2024}

\title{Complex Style Image Transformations for Domain Generalization in Medical Images}
\author{Nikolaos Spanos, Anastasios Arsenos, Paraskevi-Antonia Theofilou,\\
Paraskevi Tzouveli, Athanasios Voulodimos, Stefanos Kollias\\
National Technical University of Athens, Greece\\
{\tt\small \{nspanos,partheofilou,anarsenos\}@ails.ece.ntua.gr},\\ 
{\tt\small tpar@image.ece.ntua.gr, thanosv@mail.ntua.gr, stefanos@cs.ntua.gr}}

\begin{document}
\maketitle
\begin{abstract}
   The absence of well-structured large datasets in medical computer vision results in decreased performance of automated systems and, especially, of deep learning models. Domain generalization techniques aim to approach unknown domains from a single data source. In this paper we introduce a novel framework, named CompStyle, which leverages style transfer and adversarial training, along with high-level input complexity augmentation to effectively expand the domain space and address unknown distributions. State-of-the-art style transfer methods depend on the existence of sub-domains within the source dataset. However, this can lead to an inherent dataset bias in the image creation. Input-level augmentation can provide a solution to this problem by widening the domain space in the source dataset and boost performance on out-of-domain distributions. We provide results from experiments on semantic segmentation on prostate data and corruption robustness on cardiac data which demonstrate the effectiveness of our approach. Our method increases performance in both tasks, without added cost to training time or resources.
\end{abstract} 
\vspace{-\baselineskip}
\section{Introduction}
\label{sec:intro}

Artificial intelligence and, more specifically, deep learning and neural networks, have been at the forefront of computer vision these past years. These models generally depend on large amounts of annotated data to be able to learn properly how to extract the required information. However, there are two main issues with sufficiently training deep neural networks. The first is that for several computer vision applications, such as medical image segmentation, available data is scarce or costly to acquire, thus leading to the creation of sub-optimally trained systems. The second  is the domain shift problem \cite{zhang2021adaptive}. Deep learning systems have been shown that despite their adequate performance on data similar to the source domain, performance drops drastically when tested on data from different distributions. The first problem exacerbates the second in a large degree, thus creating the need for advanced methods to help models generalize properly to unseen domains.

Single-source domain generalization aims to address this challenge by developing robust models capable of generalizing from a single source domain \cite{ouyang2022causality}, \cite{arsenos2024uncertaintyguided}, \cite{lee2022CVPR}. By learning representations that are invariant to domain shifts and effectively disentangling domain-specific variations from task-related features, these models hold promise for improving the adaptability and performance of computer vision systems across diverse domains.

Recent methods for medical image domain generalization, such as \cite{chen2022maxstyle}, propose that creating hard image examples from a single-source dataset can be used to reliably approach domains that the model will under-perform and, thus, generalize better through failure. However, these methods requires a lot of time to train and resources, thus making it more difficult to optimize and the training method might suffer from dataset bias, since the generated images are based only on the single-source dataset.

Inspired by \cite{chen2022maxstyle}, we aim to raise performance by creating a composite data augmentation framework that leverages both feature-level space augmentation and input-level space augmentation. Through combined training and by leveraging the high-complexity input augmentation and style transfer techniques, we manage to increase the domain space so that the model can reliably generalize on unseen domain datasets.

The main contributions are:
\begin{itemize}
    \item We propose a novel framework that uses adversarial data augmentation and style transfer for domain
    generalization, while optimizing model robustness and sufficient domain space generation through complex input-level image mixing.
    \item Our method can raise out-of-domain performance, without requiring additional training time or resources.
    \item We validate our framework's performance on two different tasks for medical image segmentation and compare with the state-of-the-art.
\end{itemize}

\section{Related Work}
\label{sec:formatting}
\begin{itemize}[label={},leftmargin=*]
    \item \textbf{Deep Learning In Medical Analysis}: In recent years, the field of deep learning in medical analysis has seen remarkable advancements, especially accelerated due to the COVID-19 pandemic \cite{kollias2023ai}, \cite{arsenos2023data}, \cite{kollias2023deep}, \cite{kollias2022ai}, \cite{arsenos2022large}, \cite{kollias2021mia}, \cite{kollias2020deep}, \cite{kollias2020transparent}, \cite{gerogiannis2024covid19}, owing to the heightened global interest and the substantial accumulation of medical data. More recently, cutting-edge deep learning models are being leveraged to address a myriad of challenges in the medical domain, including image artifact correction \cite{xie2023dense}, \cite{jiang2023multiscale}, \cite{10.1007/978-3-031-43907-0_23} and text-guided applications \cite{jinginter}, \cite{zhangtpro}. We mainly focus on alleviating the domain generalization problem in the medical field using models that can be used for practical applications in medical settings, as guidance or consultation tools for doctors during clinical procedures. \\
    \item \textbf{Data Augmentation}: Data augmentation is a classic method of augmenting the training dataset to help a model generalize better. Common methods include rotating, contrast changing and cropping input images \cite{8995481}, \cite{Chen2020}, \cite{Otlora2019StainingIF}, \cite{volpi2019addressing}. More advanced methods include mixing images of different labels or replacing parts of an image with another \cite{zhang2018mixup}, \cite{yun2019cutmix}. More recent methods aim to appropriately enlarge the domain space by combining different augmentation techniques or by introducing more severe complexity in the input data, in the form of image mixing \cite{hendrycks2022pixmix}. Despite their simplicity, they are still prominent in the field of computer vision and provide a useful way of alleviating low-data performance degradation and should always be considered when creating application systems.  \\
   
    \item \textbf{Style Transfer}: Style transfer is a prominent technique in computer vision where the style of one image is enforced on another, while preserving the semantic content \cite{gatys2015neural}, \cite{park2020contrastive}, \cite{karras2019stylebased}. State-of-the-art work mainly focuses on diffusion models and their ability to extract style information from text prompts or use probabilistic functions to create new styles \cite{Zhang2023CVPR}, \cite{Wang2023ICCV}, \cite{Lu2023CVPR}. There exists heavy correlation between domain shift and alterations in image style across various domains, which can be mitigated by augmenting the diversity of training image styles \cite{dumoulin2017learned}, \cite{huang2017arbitrary}. Namely successful implementations are MixStyle \cite{zhou2021domain}, which generates ‘novel’ styles by linearly mixing style statistics from two arbitrary training instances from the same domain at feature level. DSU \cite{li2022uncertainty} perturbs feature styles with random noise to account for potential style shifts. MaxStyle \cite{chen2022maxstyle} utilizes adversarial training to generate as many negative examples of images as possible to maximize the networks robustness through constant optimizing . We utilize the MaxStyle augmentation network, combined with input level augmentations, to approach a larger domain space and create more complex and varied images.
\end{itemize}

\section{Methodology}
\label{sec:formatting}
The main objective of the data augmentation framework is to maximize robustness in a model \textit{M} from a single-source domain \textit{D}. During training, we want to create a wide domain space \textit{S} so that we can approximate various distributions \textit{\{S1,S2,\ldots\}} to simulate the domain shift. In order to achieve that, we create an augmentation pipeline that utilizes input and feature complexity to create new domains.

\begin{figure}[b]
  \centering
  \includegraphics[width=\columnwidth]{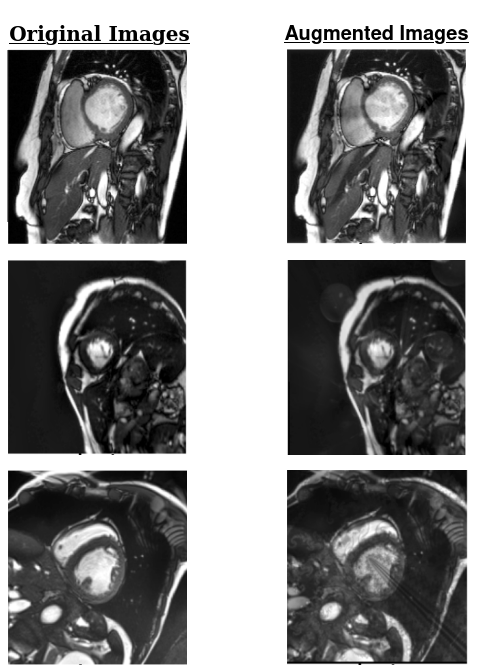} 
  \caption{\textit{Examples of original and augmented images produced by the pipeline on cardiac data.}}
  \label{fig:fig1}
\end{figure}

\subsection{Image Mixing}
We hypothesize that for enriching the domain space generated by the augmentation, not only do we require feature-level perturbations, but also input-level. Feature-level perturbations are highly reliant on the dataset used, so, despite the fact that the model is trained so that the domain space is widely generalized, it is still restricted by the input domain. Introducing images of high enough complexity during training can help not only widen the domain space generated, but also retroactively help the feature-level augmentation network create more complex images \cite{hendrycks2022pixmix}.
For the image mixing process we use fractal images and standard augmentation techniques. It has been shown in extended research that utilizing complex, non-natural images for image mixing has been beneficial for improving model performance \cite{nakashima2021vision}, \cite{kataoka2021pretraining}. We can also create fractal images using various algorithms, meaning that its relatively easy to enrich the mixing set at any time.

\begin{figure*}[t]
  \centering
  \includegraphics[width=0.8\textwidth]{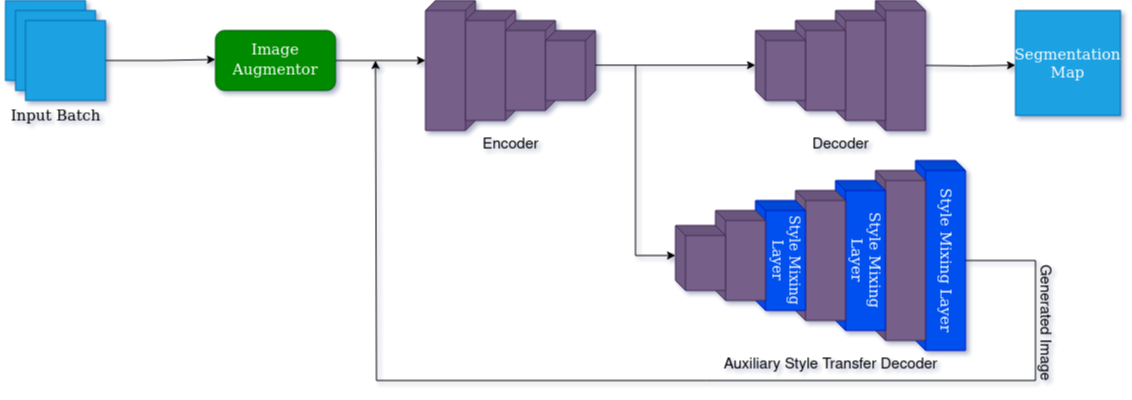} 
  \caption{\textit{The complete network architecture. Images are augmented before they enter the encoder portion of the network and the latent representations are used to create the adversarial images. Since the entire batch is mixed, the fractal images are mixed together through style mixing in the added decoder to create more varied styles during generation. Adversarial images are fed back to the training loop, while closing the auxiliary decoder, to enhance training.}}
  \label{fig:full_width}
\end{figure*}

For our framework, we randomly mix an input batch \textit{b} with fractal images, along with randomly augmenting the input batch with different, traditional augmentations \textit{t}. So, a batch after mixing is characterized by:
$b^+ = mix(t(b))$, where \textit{mix} can be additive or multiplicative functions between the fractal image and the image slice. We empirically found through our experiments that mixing intensity should not be too high, so that semantic content can be preserved and that the augmentation should not be applied across the entire training and only through controlled settings.  Otherwise, the adversarial training will lead to the creation of highly corrupted images and the backpropagation algorithm will echo the effect. We present some image examples in Figure \ref{fig:fig1}.
\subsection{Style Transfer Network}
For the style transfer network we use the method referred to in \cite{chen2022maxstyle}. The style transfer network uses an auxiliary image decoder to create adversarial samples from the latent representations of the source domain. 
Specifically, given feature \( f_i \) extracted at a certain CNN layer in the image decoder \( D_{\phi_i} \) with \( x_i \) as input, the network augments \( f_i \) via:
\begin{align}
    \text{Net}(f_i) &= (\gamma_{\text{mix}} + \Sigma_\gamma \cdot \epsilon_{\gamma}) \odot f_i \notag \\
    &\quad + (\beta_{\text{mix}} + \Sigma_\beta \cdot \epsilon_{\beta}),
\end{align}
where \( \odot \) represents element-wise multiplication. In this equation, \( f_i \) undergoes augmentation with mixed styles \( \gamma_{\text{mix}}, \beta_{\text{mix}} \), accompanied by additional style noise \( \Sigma_\gamma \cdot \epsilon_{\gamma}, \Sigma_\beta \cdot \epsilon_{\beta} \in \mathbb{R}^c \). These noise terms are introduced to facilitate exploration of potential unknown domain shifts. Style noises are sampled from a re-scaled Gaussian distribution with variances \( \Sigma_\gamma, \Sigma_\beta \in \mathbb{R}^c \), estimated from style statistics of a batch of \( B \) instances (including \( x_i \)):
\begin{align}
    \Sigma_\gamma &= \sigma^2 \left(\{\sigma(f_j)\}_{j=1,i,\ldots,B}\right), \\
    \Sigma_\beta &= \sigma^2 \left(\{\mu(f_j)\}_{j=1,i,\ldots,B}\right),
\end{align}
where $\epsilon_{\gamma}, \epsilon_{\beta} \sim \mathcal{N}(0, 1)$.

Because the style transfer network depends on the existence of sub-domains inside of a single-source domain and their subsequent mixing \cite{zhou2021domain}, this may lead to an unintended bias to the source domain dataset. Enhancing the images with high-complexity augmentation can help create more mixed domains during training at an input level, thus enhancing the model's generalization.

\subsection{Network Architecture}
We present our complete network architecture in Figure \ref{fig:full_width}. For our experiments, we utilized two different Fully Convolutional Networks (FCNs). The architecture features an encoder-decoder structure, wherein the encoder section comprises multiple convolutional layers responsible for hierarchical feature extraction from input images. These layers progressively capture both low-level and high-level representations, culminating in a condensed, lower-dimensional representation of the input features. The decoder section, on the other hand, reconstructs the spatial information from the encoded features. This allows for the generation of pixel-wise segmentation maps corresponding to the input images.

The style transfer network functions as an additional decoder that receives the latent representations as input. The noise augmentation layers are placed within the decoder layers to facilitate the style transfer process. We utilize two architectures: an FCN-16 and an FCN-64 \cite{long2015fully} for prostate data image segmentation. This investigation aims to determine the potential benefits of input augmentation techniques on the complexity of the model. Additionally, we perform evaluation on the model's robustness against image corruptions, specifically focusing on the FCN-16 architecture applied to cardiac data.

\begin{figure*}[t]
  \centering
  \includegraphics[width=\textwidth]{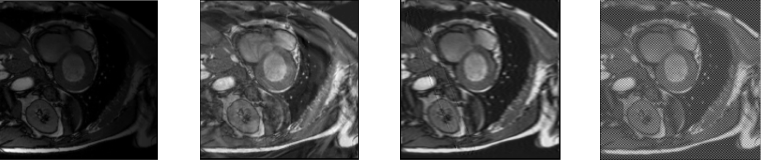} 
  \caption{\textit{Artifacted cardiac images. From left to right: Bias Artifact (showing as difference in brightness on different images areas), Ghosting Artifact (showing as idols due to movement of the organ or patient), Motion Artifact (showing as displacement), Spike Artifact (showing as stripes due to electromagnetic interference)}}
  \label{fig:fig2}
\end{figure*}


\section{Experiments and Results}
\subsection{Datasets}
We evaluate the effect of out-of-domain performance on prostate data and corruption robustness on cardiac data.
\subsubsection{Prostate Data}
For our experiments on prostate data, we utilize the publicly available prostate MRI segmentation dataset from the Medical Decathlon Challenge, which includes, in total, 148 patients.
This includes the following 7 different datasets:

\textbf{NCI-ISBI-2013}: Two different datasets from the 2013 competition, conducted by the National Cancer Institute (NCI) in collaboration with the International Society for Biomedical Imaging (ISBI). Half of the images were acquired on a 1.5T (Philips Achieva) with an endo-rectal coil (from Boston Medical Center), and the other half on a 3T (Siemens TIM) with a surface coil (from Radboud University Medical Center, Nijmegen, the Netherlands) \cite{liu2020msnet}, \cite{liu2020shapeaware}, \cite{cancerimagingarchiveNCIISBI2013}. Thus, the images are divided according to their origin for testing on out-of-domain performance. The datasets notated on the results tables are \textbf{A}, \textbf{B}.

\begin{table*}[!h]
\centering
\resizebox{\textwidth}{!}{%
\begin{tabular}{lccccccccc}
\hline
\multirow{2}{*}{Methods} & \multicolumn{7}{c}{OOD} &  & IID \\ \cline{2-8} \cline{10-10} 
 & A & B & C & D & E & F & \textbf{Mean} &  & G \\ \hline
Baseline & 87.9 & 75.5 & 77.7 & 76.8 & 29.5 & 24.2 & 58.6 &  & 83.8 \\ \hline
RandConv & 89 & 75.7 & \underline{82.3} & 85.1 & 41.4 & 52.7 & 71.03 &  & \underline{86.8} \\
AdvNoise & 87.3 & 66.4 & 76.5 & 72.8 & 44.6 & 50.3 & 66.3 &  & 82.1 \\
AdvBias & \textbf{90.2} & \textbf{84.3} & 81.1 & \textbf{87.2} & 39.7 & 51.6 & 72.35 &  & \textbf{87.4} \\ \hline
MixStyle & 87.5 & 69.3 & 80.0 & 73.4 & 43.1 & 40.8 & 65.86 &  & 84.5 \\
DSU & 88.2 & 54.4 & 81.5 & 84.1 & \underline{58.6} & 64.0 & 71.8 &  & 83.1 \\
MaxStyle & \underline{89.7} & 81.9 & 82.2 & 83.4 & 52.8 & \textbf{71.5} & \underline{76.92} &  & 85.5 \\ \hline
CompStyle (Ours) & 89.6 & \underline{83.9} & \textbf{82.4} & \underline{85.4} & \textbf{64.6} & \underline{64.5} & \textbf{78.4} &  & 86.4 \\ \hline
\end{tabular}%
}
\caption{Comparison of methods for domain generalization on prostate data with FCN-16 architecture}
\label{tab:my-table1}
\end{table*}

\begin{table*}[h]
\centering
\resizebox{\textwidth}{!}{%
\begin{tabular}{lccccccccc}
\hline
\multirow{2}{*}{Methods} & \multicolumn{7}{c}{OOD} &  & IID \\ \cline{2-8} \cline{10-10} 
 & A & B & C & D & E & F & \textbf{Mean} &  & G \\ \hline
Baseline & 90.4 & 67.9 & 79.3 & 79.1 & 45.7 & 60.9 & 70.5 &  & 83.8 \\ \hline
RandConv & 90.7 & 80.7 & 77.1 & \underline{88.0} & 27.5 & 42.2 & 67.7 &  & 87.4 \\
AdvNoise & 90.0 & 76.0 & 81.2 & 80.5 & 41.2 & 47.6 & 69.42 &  & 86.9 \\
AdvBias & \textbf{91.3} & \underline{81.6} & 82.3 & 87.2 & 45.1 & 57.5 & \underline{74.0} &  & \textbf{89.3} \\ \hline
MixStyle & \underline{91.1} & 68.9 & 82.6 & 82.4 & 43.8 & 55.4 & 70.7 &  & 85.6 \\
DSU & 90.4 & 63.6 & 79.9 & 78.2 & \underline{57.2} & \underline{61.2} & 71.75 &  & 87.3 \\
MaxStyle & 90.5 & 80.4 & \underline{84.5} & 86.0 & 45.8 & 56.0 & 73.87 &  & \underline{88.0} \\ \hline
CompStyle (Ours) & 90.7 & \textbf{86.2} & \textbf{85.7} & \textbf{88.5} & \textbf{75.2} & \textbf{77.2} & \textbf{83.92} &  & 86.4 \\ \hline
\end{tabular}%
}
\caption{Comparison of methods for domain generalization on prostate data with FCN-64 architecture}
\label{tab:my-table2}
\end{table*}

\begin{table*}[!h]
\centering
\resizebox{\textwidth}{!}{%
\begin{tabular}{@{}cccccccccccccccc@{}}
\hline
\multirow{3}{*}{Methods} & \multicolumn{15}{c}{Corruption}                                                                                                           \\ \cmidrule(l){2-16} 
                         & \multicolumn{3}{c}{RandBias} &  & \multicolumn{3}{c}{RandSpike} &  & \multicolumn{3}{c}{RandMotion} &  & \multicolumn{3}{c}{RandGhosting} \\ \cmidrule(lr){2-4} \cmidrule(lr){6-8} \cmidrule(lr){10-12} \cmidrule(l){14-16} 
                         & LV       & MYO     & RV      &  & LV       & MYO      & RV      &  & LV       & MYO      & RV       &  & LV        & MYO       & RV       \\ \midrule
Baseline                 & 91.9     & 85.5    & 83.2    &  & 1        & 1        & 1       &  & 91.6     & 83.6     & 82.8     &  & 91.1      & 83.7      & 81.6     \\
MaxStyle                 & 91.5     & 84.9    & 83.2    &  & 57.9     & 48.4     & 44.9    &  & 91.8     & 84.2     & 83.9     &  & 91.0      & 83.6      & 82.7     \\
CompStyle (Ours)        & 91.9     & 84.8    & 83.4    &  & 62.5     & 49.5     & 51.0    &  & 91.6     & 83.9     & 83.0     &  & 90.9      & 83.4      & 80.3     \\ \bottomrule
\end{tabular}%
}
\caption{Comparison table of methods for domain generalization  on Cardiac Data with FCN-16 Architecture}
\label{tab:my-table3}
\end{table*}

\textbf{I2CVB}: A dataset from the Initiative for Collaborative Computer Vision Benchmarking (I2CVB) \cite{liu2020msnet}, \cite{liu2020shapeaware}, \cite{lemaitre2015computer}. The images were taken by a 3T scanner Siemens, with the following methods for better representation: Magnetic Resonance Imaging with T2 weighting (T2-W), Dynamic Contrast-Enhanced Magnetic Resonance Imaging (DCE), Diffusion-Weighted Magnetic Resonance Imaging (DWI), Magnetic Resonance Spectroscopic Imaging (MRSI). The dataset notated on the results tables is \textbf{C}.

\textbf{PROMISE12}: Three datasets from different medical centers and different acquisition methods from the PROMISE12 competition for prostate medical image segmentation \cite{liu2020msnet}, \cite{liu2020shapeaware}, \cite{grandchallengePROMISE12Grand}. The datasets notated on the results tables are \textbf{D}, \textbf{E} and \textbf{F}.

\textbf{Medical Decathlon Dataset}: A new dataset provided by the challenge competition. This specific dataset was used for training and validation in within-distribution applications \cite{antonelli2021medical}. The dataset notated on the results tables is \textbf{G}.

\subsubsection{Cardiac Data}
For our experiments we evaluate corruption robustness on the publicly available dataset the ACDC cardiac segmentation challenge dataset \cite{8360453}, consisting of a total of 100 heart MRI scans. Of the three-dimensional images, we make use of the slices corresponding to end-diastole and end-systole periods. The dataset is multi-labelled, consisting of annotations for left ventricle, myocardium and right ventricle. We evaluate on the following 4 types of corruptions created by TorchIO software \cite{PrezGarca2021}:
\begin{itemize}[leftmargin=*]
    \item Motion: Random "motion", which usually appears physiologically in MRI images due to organ motion.
    \item Spike: Random spike artifacts, also known as Herringbone artifacts, create stripes in different directions in the image space due to spikes generated in the machine's electromagnetic field.
    \item Ghosting: Random displacement of an image ghost, usually due to cardiac motion, patient motion during the examination, or blood flow.
    \item Bias Field: Random fluctuations in image intensity usually due to the in-homogeneity of the MRI machine's field.
\end{itemize}
Examples of the images are shown in Figure \ref{fig:fig2}.

\subsection{Implementation and experiment set-up}
We follow the same training schema and dataset preprocessing as in \cite{chen2022maxstyle}. For the style transfer layers of the dual-branch network, we raise the number of iterations required for optimizing the adversarial images. For the input-level augmentation, we lower the intensity from the default parameters to ensure preservation of semantic content in the input images. Each batch in training is either mixed by the augmentation with a probability of 0.4 or noise is added to prevent overfitting. We evaluate the composite augmentation frameworks on two different architectures, FCN-16 and FCN-64. We retain the Adam optimizer and multi-task loss from \cite{chen2022maxstyle} and train for 600 epochs with batch size 20 on the FCN-16 architecture for both tasks and for 600 epochs with batch size 15 on the FCN-64 architecture. All experiments were conducted on an NVIDIA V100 GPU and an NVIDIA GeForce 4060Ti GPU. We compare the augmentation frameworks with competitive methods such as: RandConv \cite{xu2021robust}, adversarial noise \cite{miyato2018virtual}, adversarial bias field \cite{chen2020realistic}, MixStyle \cite{zhou2021domain}, DSU \cite{li2022uncertainty} and MaxStyle \cite{chen2022maxstyle}. We use the DICE score metric for evaluation, averaged on the entirety of each dataset \cite{ronneberger2015unet}. All experiments were re-done locally due to an error in the evaluation of \cite{chen2022maxstyle}, which was noted in their code implementation.

\subsection{Results}
Results for the prostate segmentation task for the FCN-16 and FCN-64 architectures can be viewed on tables \ref{tab:my-table1} and \ref{tab:my-table2} respectively. We present the best result for evaluation from our framework. Input-level methods generally achieved better intra-domain performance than the style-transfer feature-level methods, especially AdvBias. However, both MaxStyle and DSU outperformed the rest on the \textbf{E} and \textbf{F} datasets. We can see from the results that the composite augmentation framework achieves the best out-of-domain performance in both architectures. Increasing the amount of features within the architecture help boost performance for all datasets on the composite framework, while the other methods suffered heavily from the added complexity. Intra-domain performance for our framework remains competitive with the rest of the methods, albeit with a small drop in scoring. The small drop may be impacted by the large number of unknown domains introduced during training. Other methods can be more favorable towards the source domain, thus appearing in evaluation as performing better. However, favoring the source domain means that out-of-domain performance will severely drop, which is detrimental towards developing practical systems. In general, the added input complexity combined with a larger network helped create a more stable out-of-domain performance on all datasets.

We present results for corruption robustness on table \ref{tab:my-table3}. We compare with MaxStyle since it showed the best performance on the FCN-16 architecture. Of all four artifacts, RandSpike appeared to be the most detrimental towards the model's performance, which is in-line with how much it influences the entire image. The composite augmentation helped boost performance towards the most detrimental artifact and remained competitive on all labels with the rest of the artifacts. 

\section{Conclusion}
We introduced a new augmentation framework for domain generalization from a single-source domain. It combines complex image augmentations and a style transfer adversarial network to enrich the domain space during the training process. By creating images of diverse and hard styles, we raise network robustness towards the domain shift problem and assist in boosting the model's generalization. Higher complexity networks can better utilize the added input complexity, in comparison with networks that have fewer features in their layers. We also highlight the importance of tuning the augmentation in a way that the mixing process does not alter the semantic content, especially for medical images where it is vital that the correct information is fed to the model. We validated the effect of the augmentation on two different tasks, for prostate segmentation and cardiac data corruption robustness. From our experiments, we conclude that certain methods favor certain datasets. Degradation can be task-dependent or even dataset-dependent, but utilizing enough image complexity during training can help leverage an even increase for all out-of-domain datasets. Our method works without a need for more resources or more training time, proving the importance of standard input augmentation techniques when tackling model generalization. For future work, it would be interesting to see the effect of combining other techniques, such as methods of adding noise in the encoder portion of the network, or these methods validated on more recent image segmentation networks, such as vision transformers, or different tasks, such as autonomous driving or agricultural applications.

\section{Acknowledgements}
This research work is Co-funded from the European Union’s Horizon Europe Research and Innovation programme under Grant Agreement No 101119714 — dAIry 4.0. 


{
    \small
    \bibliographystyle{ieeetr}
    \bibliography{main}
}


\end{document}